\documentclass[dvips,12pt, twoside, letterpaper]{article}
\usepackage{graphicx}
\usepackage{color}

\title{Exploring the underlying mechanisms of Xenopus laevis embryonic cell cycle}

\author{Kun Zhang$^a$ Jin Wang$^{ab*}$\\
\small{$^a$ State Key Laboratory of Electroanalytical Chemistry } \\
\small{Changchun Institute of Applied Chemistry } \\
\small{Chinese Academy of Sciences } \\
\small{Changchun, Jilin 130022 } \\
\small{People's Republic of China } \\
\small{$^b$ Department of Chemistry and Physics,}
\small{Department of Applied Mathematics,} \\
\small{State University of New York at Stony Brook } \\
\small{Stony Brook, NY 11794 } \\
\small{$^*$ Corresponding Author: E-mail:jin.wang.1@stonybrook.edu}}

\begin{document}
\maketitle

\newpage
\begin{abstract}
Cell cycle is an indispensable process in the proliferation and development. Despite significant efforts, global quantification and physical understanding are still challenging. In this study, we explored the  mechanisms of Xenopus laevis embryonic cell cycle by quantifying the underlying landscape and flux.  We uncovered the irregular Mexican hat landscape of the Xenopus laevis embryonic cell cycle with several local basins and barriers on the oscillation path. The local basins characterize the different phases of Xenopus laevis embryonic cell cycle and the local barriers represent the checkpoints. The checkpoint mechanism of cell cycle is revealed by the landscape basins and barriers. While landscape shape determines the stabilities of the states on the oscillation path, the curl flux force determines the stability of the cell cycle flow. Replication is fundamental for biology of living. From our quantitative study here, we see that replication can not proceed without energy input. In fact, the curl flux originated from energy or nutrition supply determines the speed of the cell cycle and guarantees the progression. Speed of cell cycle is a hallmark of cancer. Through landscape and flux analysis, one can identify the key elements for controlling the speed. This can help to design effective strategy for drug discovery against cancer.
\end{abstract}


\newpage

\section*{Introduction}
The cell cycle is a periodic process in the biological cell that duplicates its own components and divides into two daughter cells. In this process, the genetic material containing DNA molecule is accurately replicated, and then the two copies are separated into the daughter cells during division. Exploring the mechanism of the cell cycle is important for understanding cell growth,development, reproduction and death\cite{Chen2000-369,Tyson2001-249,Chen2004-3841,Tyson2008-R759}. The complete cell cycle is often composed of four phases: the synthesis of DNA(S phase), mitosis(S phase), and the intervening phase G1 and G2 and M phase. M phase is often divided into four subphases: prophase, metaphase, anaphase and telophase. Sometimes, the cell can enter into a state of quiescence called G0 phase in which the cell temporarily or reversibly stops dividing. To ensure the proper progression of the cell division, the cell cycle checkpoints control the ordering of the cell cycle. This leads to the starting of each phase dependent on the completion of the pervious one. It is now believed that the cell cycle process is tightly controlled by the underlying gene regulatory network. With the increasing understanding of the biology, the mathematical models have been proposed to uncover the mechanisms \cite{Tyson2001-249,Goldbeter1991-9107,Qi2013-273}. The key of these models is the activity of cyclin-dependent kinases(CDKs) and their associated cyclin protein, which jointly dominate the process of the cell cycle\cite{Nurse1998-1103}.

It is still challenging to see exactly how the underlying gene regulatory network control the cell cycle progressing because of the complexity of the network. In addition, there are intrinsic fluctuations from the finite number of molecules and extrinsic fluctuations from inhomogeneous environments in the living cells\cite{Swain2002-12795,Thattai2001-8614}. Therefore, the stochastic nature must be considered in the studying the cell cycle \cite{Huang1996-10078,Elowitz2000-335,Ideker2001-929,Davidson2002-1669,Yu2006-1600,Kar2009-6471}. Although the stochastic nature of gene regulatory network has been studied, it is still challenging to have global quantifications and physical explanations for cell cycle to reveal its underlying mechanisms.

In this work, we explore the mechanisms of Xenopus laevis embryonic cell cycle controlled by the underlying gene regulatory network. We do so by the quantifications of the underlying landscape and the flux.  \cite{Sasai2003-2374,Wang2008-12271,Ao2009-63,Wang2010-8195,Wang2011-8257,Li2013-}. The different gene expression patterns in the cell cycle can be represented in the state space of the underlying gene regulatory network. There are many such states in the state space. Not every state is equally probable. The weight of occurrence for every gene expression pattern can be described by the probability distribution in state space.  The higher probability means higher chance of appearance which can be observed in the experiments. The specific functional states or phases of the cell correspond to the specific gene expression patterns, often with higher probabilities (or lower potential valleys) on the landscape. By quantifying the topography of the potential landscape through the barrier heights and kinetics between different basins as well as the underlying curl flux, we can identify the driving force of the cell cycle and explore the global stabilities of the oscillation states and the flow of the Xenopus laevis embryonic cell cycle. Furthermore, we can quantify the energy dissipation of the cell cycle and investigate the origin of the curl flux and the speed of the Xenopus laevis embryonic cell cycle. We show that cell replication can not proceed without energy input. By exploring the relationship between the entropy production rate (energy cost) and speed of the cell cycle, we can understand this at the quantitative level. The speed of the cycle is a hallmark of cancer. Through the landscape and flux analysis, we can find out the key elements controlling the speed of the cell cycle. This can help to design effective strategy for drug discovery against cancer.

\section*{Results and Discussions}

\subsection*{Cell cycle model}
\newcommand{\ud} {\mathrm{d}}
Mathematical models of cell cycle have been proposed \cite{Chen2000-369,Tyson2001-249,Wang2010-8195,Yang2013-519,Li2014-14130,Gerard2009-21643,Gerard2010-simple,Gerard2012-Entrainment}.
In this study, we introduce a two-gene model of the underlying Xenopus laevis embryonic cell cycle network which includes fewer adjustable parameters. The simpler model is beneficial to analyze the general nature  and compare to other activator-repressor circuits. This cell cycle gene circuit was proposed earlier \cite{Yang2013-519}. The model was represented by the following two equations
\begin{eqnarray}
\frac{\ud}{\ud t}Cyc&=&k_s-k_{deg}Cyc\nonumber \\
\frac{\ud}{\ud t}Cdk1&=&k_s+k_{cdc}(Cyc-Cdk1)-k_{weel}*Cdk1-k_{deg}Cdk1 \nonumber
\end{eqnarray}
Where the first equation describes the synthesis and degradation of the moitotic cyclins. $k_s$ is a rate constant of cyclin synthesis, $k_{deg}$ is rate constant of cyclin degradation which varied with the activity of Cdk1 and can be described by Hill function:$k_{deg}=a_{deg}+b_{deg} \frac{Cdk1^{n_{deg}}}{EC50_{deg}^{n_{deg}}+Cdk1^{n_{deg}}}$. The second equation describes the production of active Cdk1. The parameter $k_{cdc}$ represents the production rate of protein Cdc25C which can activate Cdk1 by removing phosphate. The parameter $k_{weel}$ denotes the production rate of protein kinase Wee1A which can repress the active Cdk1. Both $k_{cdc}$ and $k_{weel}$ are the functions of the active Cdk1 concentration. Their steady-state response was determined by experimental studies and can be approximated by the Hill functions: $k_{cdc}=a_{cdc}+b_{cdc}\frac{Cdk1^{n_{cdc}}}{EC50_{cdc}^{n_{cdc}}+Cdk1^{n_{cdc}}}$, $k_{weel}=a_{weel}+b_{weel}\frac{EC50_{weel}^{n_{weel}}}{EC50_{weel}^{n_{weel}}+Cdk1^{n_{weel}}}$.

This two dimensional model assumed that there was no time lag between the activation of Cdk1 and regulation of cyclin degradation. Considering the realistic time lag into the negative feedback loop, the model can be expressed with the following equation:
\begin{eqnarray}
\frac{d}{dt}Cyc&=&k_s-(a_{deg}+b_{deg}APC_{30})Cyc\nonumber \\
\frac{d}{dt}Cdk1&=&k_s+(a_{cdc}+b_{cdc}\frac{Cdk1^{n_{cdc}}}{EC50_{cdc}^{n_{cdc}}+Cdk1^{n_{cdc}}}(Cyc-Cdk1)\nonumber \\
&&-(a_{weel}+b_{weel}\frac{EC50_{weel}^{n_{weel}}}{EC50_{weel}^{n_{weel}}+Cdk1^{n_{weel}}})*Cdk1-(a_{deg}+b_{deg}APC_{30})Cdk1 \nonumber \\
\frac{d}{dt}APC_0&=&-k_{p}Cdk1 APC_0+k_{d}APC_1\nonumber \\
\frac{d}{dt}APC_1&=&k_{p}Cdk1 APC_0-k_{d}APC_1 - k_{p}Cdk1 APC_1 +k_d APC_2\nonumber \\
\frac{d}{dt}APC_2&=&k_{p}Cdk1 APC_1-k_{d}APC_2 - k_{p}Cdk1 APC_2 +k_d APC_3\nonumber \\
... \nonumber \\
\frac{d}{dt}APC_{29}&=&k_{p}Cdk1 APC_{28}-k_{d}APC_{29} - c k_{p}Cdk1 APC_{29} +\frac{1}{c}k_d APC_{30}\nonumber \\
\frac{d}{dt}APC_{30}&=&c k_{p}Cdk1 APC_{29} -\frac{1}{c}k_d APC_{30} \nonumber
\end{eqnarray}
Where APC has 31 phosphorylated forms and only the final form is active. The constant c denotes the cooperativity of the phosphorylation and dephosphorylation reactions.

\subsection*{Landscape and flux of the cell cycle system}

Based on the underlying gene regulatory network, we investigate the associated stochastic dynamics. By following the probabilistic evolution, we can quantify the steady state probability distribution in the expression state space.
We found the underlying potential landscape of the limit cycle dynamics of the Xenopus laevis embryonic cell cycle has a irregular Mexican hat shape. Outside the oscillation cycle ring, the dynamics is attracted to the ring mainly by the negative landscape gradient, which guarantees the stability of the states on the oscillation path. Once on the oscillation ring, the driving force for the Xenopus laevis embryonic cell cycle is then mainly determined by the curl flux along the cycle along with the impeding force from the local basins and barriers on the cycle. We found a few local basins and barriers between the basins along the Xenopus laevis embryonic cell cycle trajectory. These local basins on the potential landscape reflect the different phases of the Xenopus laevis embryonic cell cycle processes. The saddle points between the basins on the landscape quantify the checkpoints of different stages of Xenopus laevis embryonic cell cycle. These quantifications can be used to uncover the checkpoint mechanisms of the Xenopus laevis embryonic cell cycle from physical perspectives. Through the global picture via the landscapes and flux on cell cycle progression, we can quantify the global stability and function of the Xenopus laevis embryonic cell cycle. To complete the cell cycle progression, the cell cycle process has to overcome the major barriers by sufficient driving force. In other words, the Xenopus laevis embryonic cell cycle progression must prepare adequately to pass through each checkpoint. The curl flux provides such a driving force. We can also estimate the energy dissipation required for maintaining the Xenopus laevis embryonic cell. We found it is correlated with the curl flux. Therefore,  the Xenopus laevis embryonic cell cycle process requires the energy input through nutrition supply. We further found the period and coherence of the Xenopus laevis embryonic cell cycle is strongly correlated with curl flux.

\begin{figure}[!ht]
\centering
\includegraphics[width=1.0\textwidth]{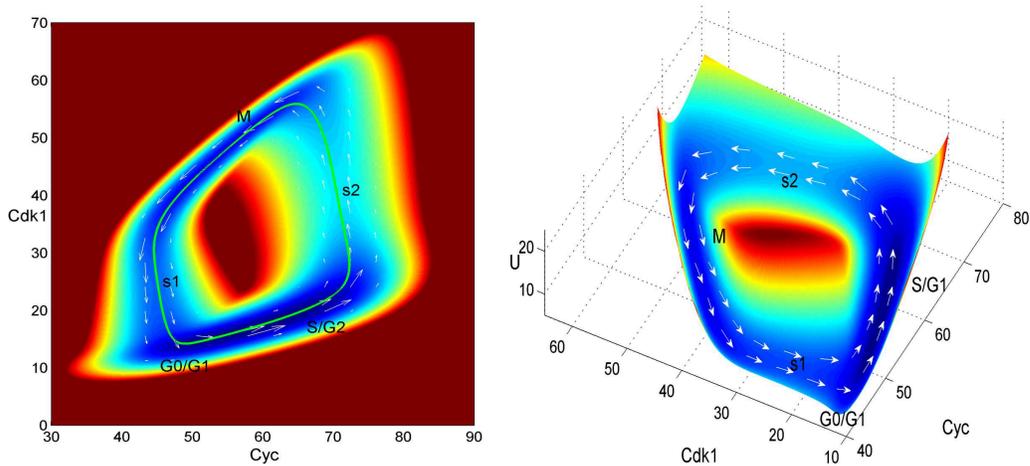}
\caption{(a)The two dimensional Landscape of 2-dimension cell cycle model;(b)The Three dimensional Landscape of 2-dimension cell cycle model;}\label{fig01}
\end{figure}

The Figure \ref{fig01}(a) shows the two dimensional potential landscape of the two variables Xenopus laevis embryonic cell cycle model. we find the landscape has two valley basins and two saddle points. The bottom basin has a long and narrow shaped valley. It represents the G0/G1 phase and S/G2 phase on each side of the valley respectively. The saddle point s2 along the cell cycle path is G2 checkpoint, which can guarantee that DNA replication is achieved before reaching to the next phase M. The top basin represents that the cell attains the M phase. When a cell matures and the division occurs, the cell goes through saddle point s1 from the phase M back to the G0/G1 phase. The saddle point s1 is M checkpoint. The Figure \ref{fig01}(b) show the three dimensional potential landscape of two variable Xenopus laevis embryonic cell cycle model. The figure further represents the progression of the cell cycle by the landscape and the flux(white arrow).

\begin{figure}[!ht]
\centering
\includegraphics[width=1.0\textwidth]{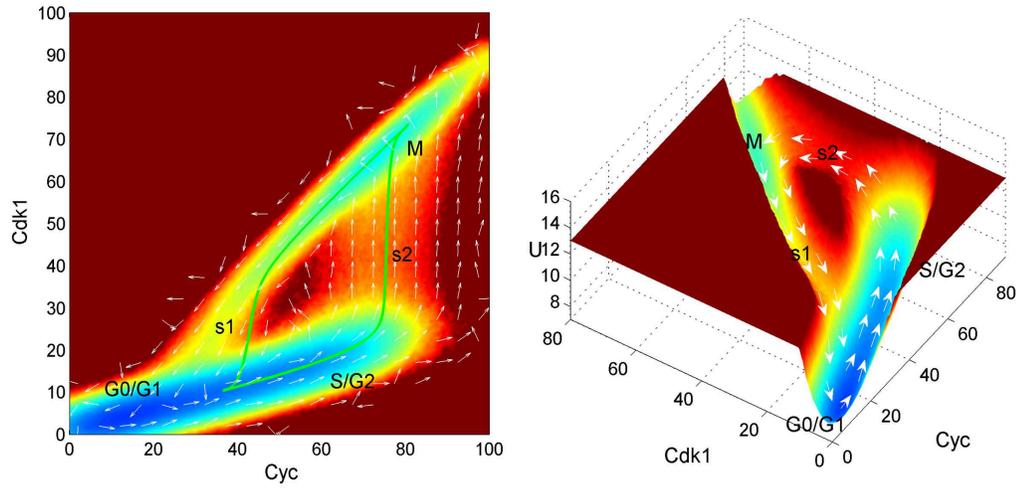}
\caption{(a)The two dimensional Landscape of 33-dimension cell cycle model;(b)The Three dimensional Landscape of 33-dimension cell cycle model;}\label{fig02}
\end{figure}

The Figure \ref{fig02}(a) show the two dimensional landscape of the more sophisticated 33 variables cell cycle model. The white arrow is the probability curl flux. The negative gradient of potential landscape attracts the system to the oscillation path. On the oscillation path, the curl flux guarantees the stable cell cycle flow. The landscapes show similar process of mitosis with both 2 variable and 33 variable model, but the total cyclin and active Cdk1 of the G0/G1 phase are lower in the 33 variable model. We also see in 33 variable model that there is a sharp increase in CDK1 from S/G2 (together with G0/G1 under this parameter range) to M phase and there is also a sharp decrease of CDK1 from M phase back to G0/G1 phase in contrast to the 2 variable model. The Figure \ref{fig02}(b) show the three dimensional potential landscape of 33 variable Xenopus laevis embryonic cell cycle model.

\begin{figure}[!ht]
\centering
\includegraphics[width=1.0\textwidth]{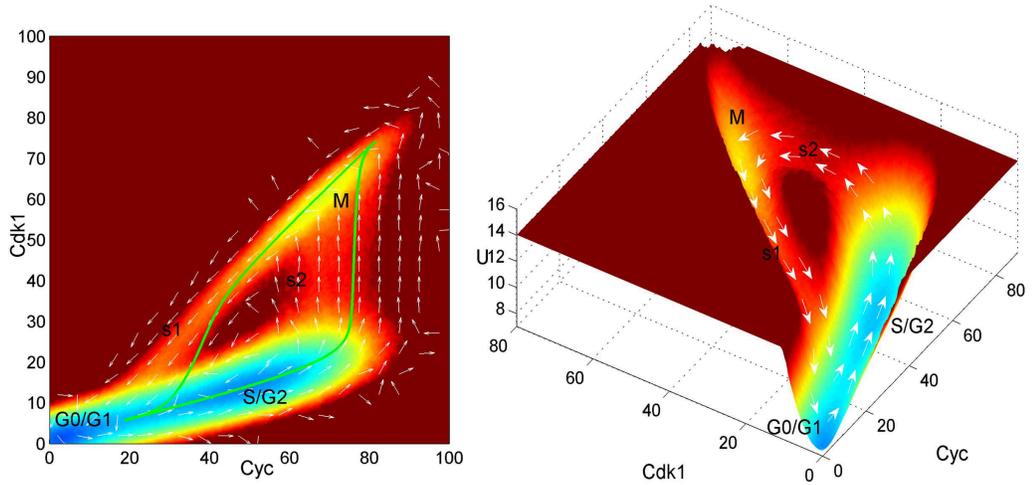}
\caption{(a)The two dimensional Landscape of 33-dimension cell cycle model with parameter $b_{deg}=0.18$;(b)The Three dimensional Landscape of 33-dimension cell cycle model with parameter $b_{deg}=0.18$;}\label{fig03}
\end{figure}

The Figure \ref{fig03}(a) and \ref{fig03}(b) show the two and three dimensional potential landscape of 33 variables Xenopus laevis embryonic cell cycle network with different parameters from the figure \ref{fig02}. In this parameter setting, the G0/G1 phase and S/G1 phase are in different attractor basins. This gives a G1 checkpoint between the G0/G1 phase and S phase.

\subsection*{The Effects of Some Parameters for Cell Cycle System}

\begin{figure}[!ht]
\centering
\includegraphics[width=1.0\textwidth]{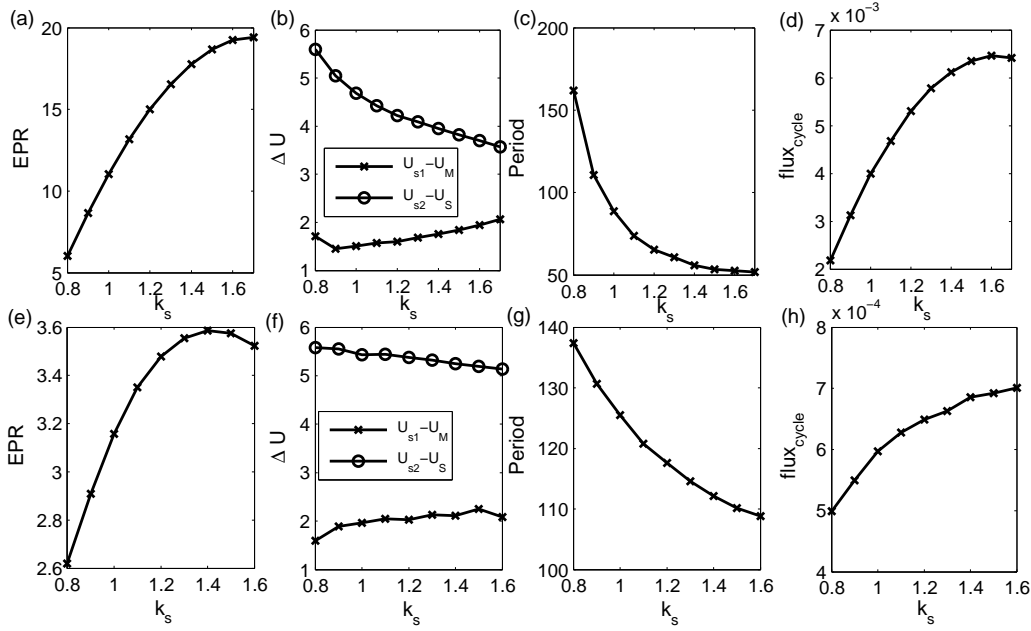}
\caption{(a)(e)Entropy production rate with different $k_s$.(b)(f)Barrier Height versus parameter $k_s$.(c)(g)The period with different $k_s$.(d)(h)The integral of flux along the limit cycle $\oint Jdl/dl$ versus parameter $k_s$. (Top)2-dimension model; (Bottom)33-dimension model;}\label{fig04}
\end{figure}

To study the energy cost of the cell cycle processes, we compute the entropy production rate with different parameters. As shown in Figure\ref{fig04}(a), the entropy production rate increases with the increase of cyclin synthesis rate. This result states that the increasing of cyclin synthesis rate requires more energy or nutrition supply. In Figure \ref{fig04}(b), we show the changes of the two barrier heights with different cyclin synthesis rate constants. One is the barrier height between the saddle s2 and the bottom basin. It characterizes the G2 checkpoint and quantifies the degree of difficulty of the cell cycle from G2 phase to M phase. The other is the barrier height between the saddle s1 and the top basin. It denotes the degree of difficulty of the cell cycle from M phase back to G0/G1 phase. We find the first barrier through s1 decreases with the increasing of cyclin synthesis rates. It indicates the division of a cell is likely to be easier under this condition. The latter barrier increases with the increasing of synthesis rate. It states that a cell in the S/G2 phase becomes more difficult to get to the M phase through the G2 check point. In Figure \ref{fig04}(c), we calculated the period of cell cycle and find the period decreases with the increasing of cyclin synthesis rates. The period can reflect how fast the cell cycle oscillates and its growth rate. So the result implies that the increasing of cyclin synthesis rate can accelerate the cell cycle and its growth. In Figure \ref{fig04}(d), we computed the integral of flux along the limit cycle with different cyclin synthesis rate. We find the integral of flux increases with the cyclin synthesis rate. This reflects that the flux strengthens upon the increasing of synthesis rate. The flux and the entropy production rate have the same tendency with respect to cyclin synthesis rates. So the flux of cell cycle is closely related to the entropy production rate. With the increasing of cyclin synthesis rate, the Xenopus laevis embryonic cell cycle has the stronger flux and higher entropy production rate. This will consume more the energy or nutrition supply. Meanwhile, the cell cycle from the maturation to the division becomes faster. It is worthwhile to notice that the barrier height and flux together determine the dynamics of the cell cycle such as cycle speed or period. Even though the barrier is higher between S/G2 to M phase, the flux can still drive the cell cycle through the s2 saddle or transition state (G2 check point) with faster speed\cite{Dart2004-16433,Xu2012-,Zhang2013-14930}. Similar conclusions can be drawn from the corresponding the more sophisticated 33 variable network.

\begin{figure}[!ht]
\centering
\includegraphics[width=1.0\textwidth]{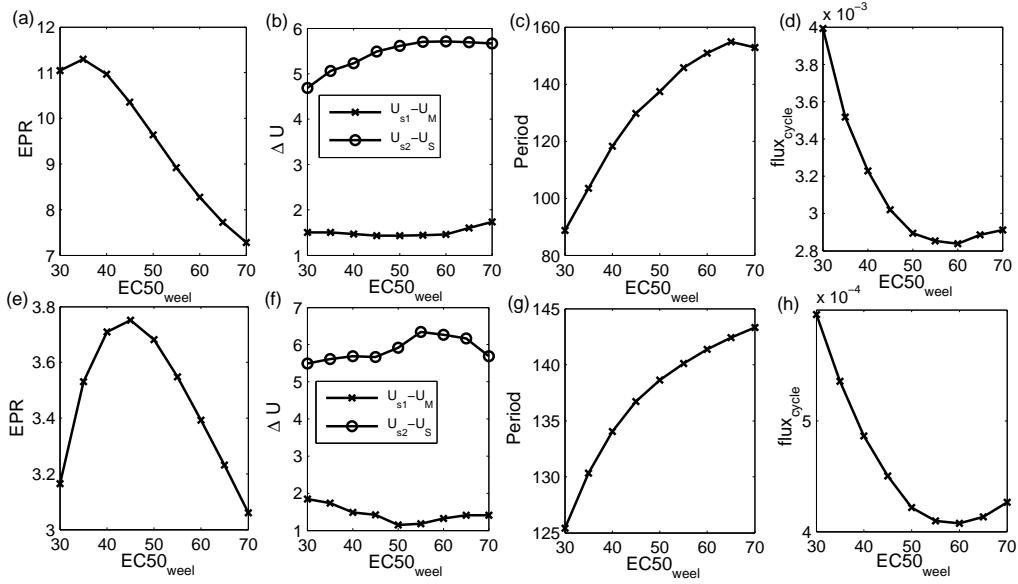}
\caption{(a)(e)Entropy production rate with different $EC50_{Weel}$.(b)(f)Barrier Height versus parameter $EC50_{weel}$.(c)(g)The period with different $EC50_{Weel}$.(d)(h)The integral of flux along the limit cycle $\oint Jdl/dl$ versus parameter $EC50_{Weel}$ (Top)2-dimension model; (Bottom)33-dimension model;}\label{fig05}
\end{figure}

As shown in Figure \ref{fig05}(a), the entropy production rate decreases with the increasing of parameter $EC50_{Weel}$. The parameter is Hill constant or the half-maximum effective concentration values of Wee1A as the substrate of Cdk1. This results state that the increasing of parameter $EC50_{Weel}$ leads to less energy cost. In Figure \ref{fig05}(b), we show the change of two barrier heights with different parameters in $EC50_{Weel}$. One is the barrier height between the saddle s2 and the bottom basin. The other is the barrier height between the saddle s1 and the top basin. They denote the degree of the difficulty in cell cycle from current phase to the next phase. We find that the barriers change smoothly with the increasing of parameter $EC50_{Weel}$ . It indicates that changes in this parameter have moderate impact on the growth and division of a cell under these conditions. In Figure \ref{fig05}(c), we calculated the period of cell cycle and find that the period increases with the increasing of this parameter. Therefore, the trend in cell cycle period shows that the increasing of the parameter $EC50_{Weel}$ can decelerate the cell cycle and its growth. In Figure \ref{fig05}(d), we computed the integral of the curl flux along the Xenopus laevis embryonic cell cycle with different parameters in $EC50_{Weel}$. We find that the integral of flux declines with parameter $EC50_{Weel}$. Similarly, the flux and the entropy production rate have the same tendency under the changes of the parameter $EC50_{Weel}$. Similar conclusions can be drawn from 33 variable model of the Xenopus laevis embryonic cell cycle.

\subsection*{The Coherence for Cell Cycle System}

\begin{figure}[!ht]
\centering
\includegraphics[width=1.0\textwidth]{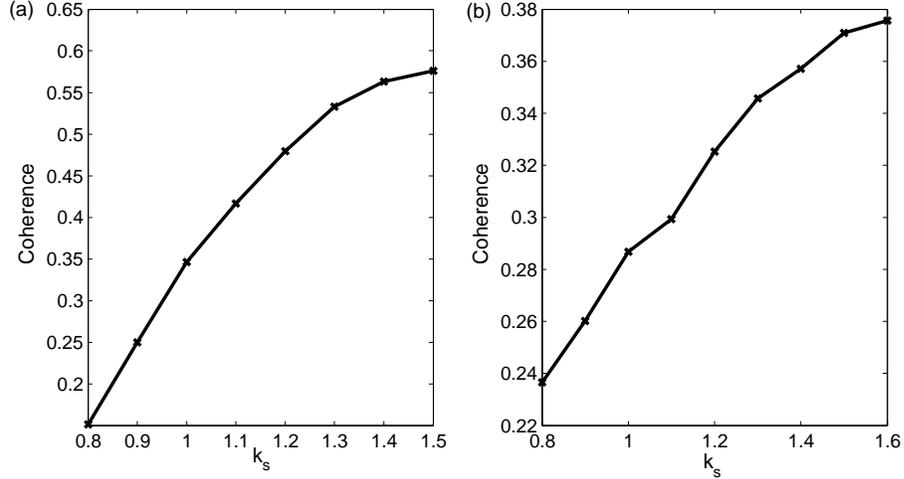}
\caption{(a)The coherence versus parameter $k_s$ with 2d model(b)The coherence versus parameter $k_s$ with 33d model}\label{fig08}
\end{figure}

The coherence can quantify the stability of the oscillation and measure the degree of persistence of the oscillatory phase. In Figure \ref{fig08}, we plot the relationship between the cyclin synthesis rate and the coherence. we find that the coherence increases with the increasing of synthesis rate. Meanwhile, we know the flux also becomes stronger with the increasing of the synthesis rate. This states that the enhancement of the flux can improve the stability and coherence of the periodic oscillation and persistence the cell cycle.

\begin{figure}[!ht]
\centering
\includegraphics[width=1.0\textwidth]{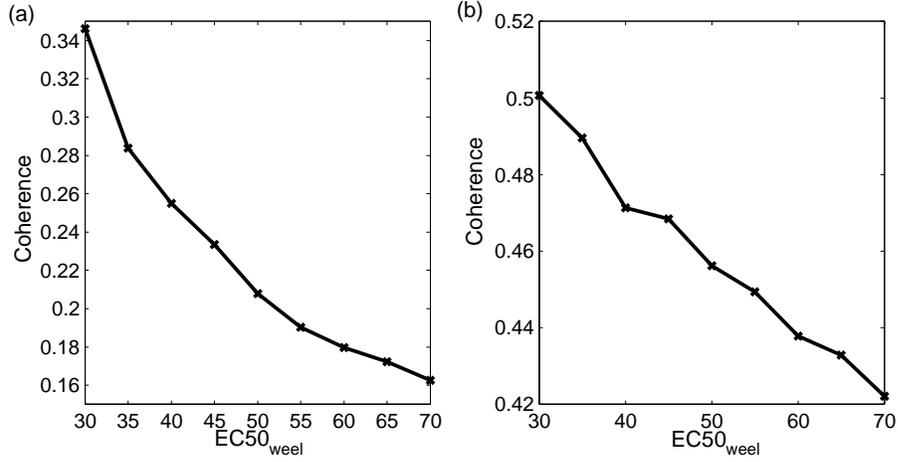}
\caption{(a)The coherence versus parameter $EC50_{weel}$ with 2d model (b)The coherence versus parameter $EC50_{weel}$ with 33d model}\label{fig09}
\end{figure}

In Figure \ref{fig09}, we draw the change of the coherence with the parameter $EC50_{Weel}$. we find that the coherence declines with the increasing of parameter $EC50_{Weel}$. Meanwhile, the flux also becomes weaker with the increasing of this parameter. It implies that less flux can decrease the stability of the oscillation flow and reduce the cell cycle speed. Flux is thus crucial for maintaining the cell cycle.

\begin{figure}[!ht]
\centering
\includegraphics[width=1.0\textwidth]{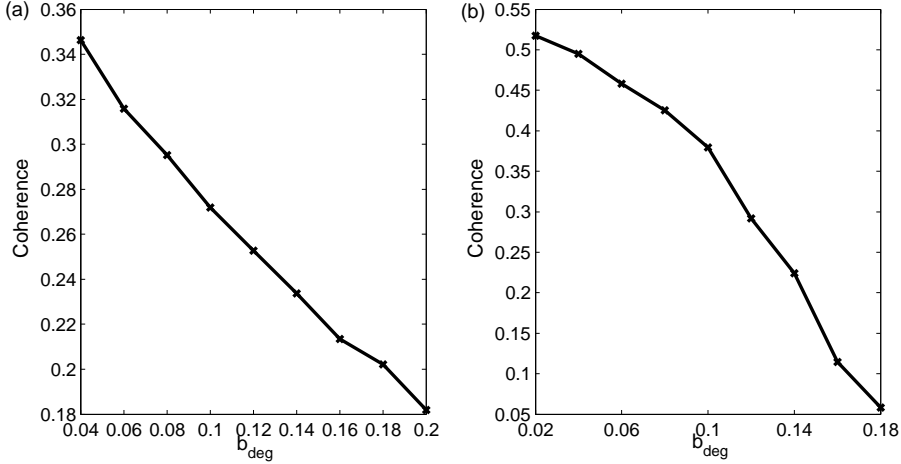}
\caption{(a)The coherence versus parameter $b_{deg}$ with 2d model(b)The coherence versus parameter $b_{deg}$ with 33d model}\label{fig10}
\end{figure}

In Figure \ref{fig10}, we draw the change of the coherence with the degradation parameter $b_{deg}$. we find that the coherence declines with the increasing of parameter $b_{deg}$. Meanwhile, the flux also become weaker with the increasing of the degradation parameter$b_{deg}$ . Flux is important for the stability of the oscillation flow and the maintaining of the cell cycle.

\subsection*{The Relationship among The Entropy Production Rate, The Flux and The Period}

\begin{figure}[!ht]
\centering
\includegraphics[width=1.0\textwidth]{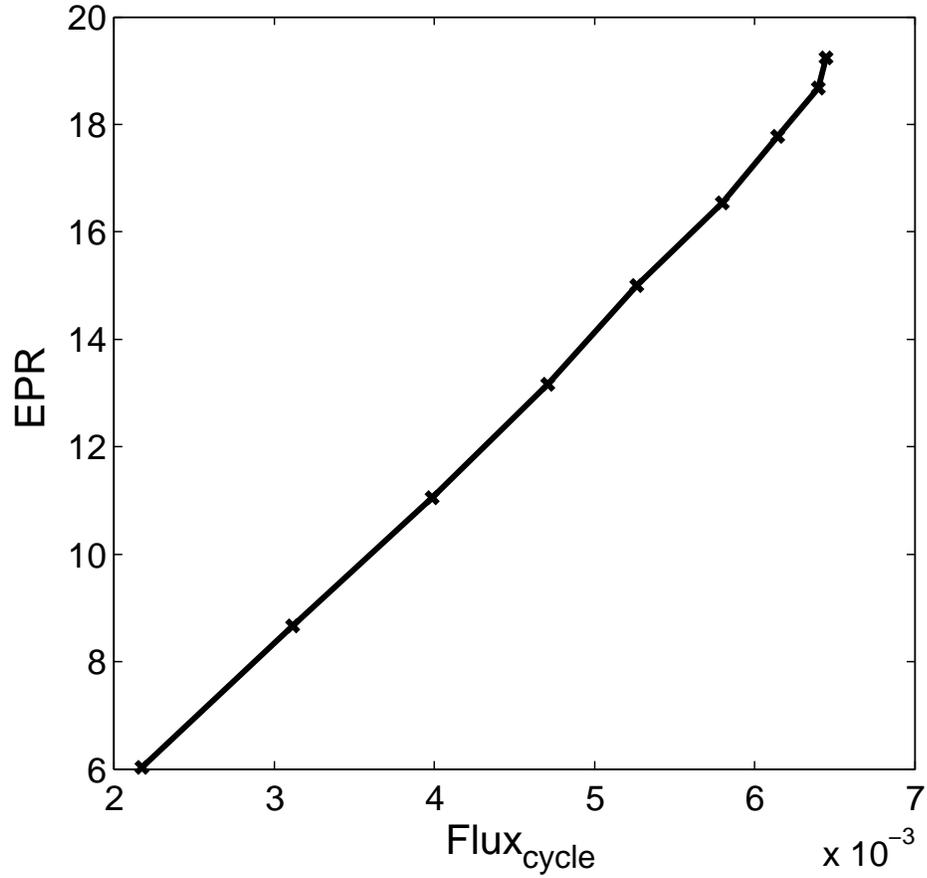}
\caption{The Entropy production rate versus the integral of flux along the limit cycle}\label{fig11}
\end{figure}

In Figure \ref{fig11}, we show the change of EPR with the flux by an example of parameter $k_{s}$. We find the EPR increase with the increasing of the flux of limit cycle. This implies the cost energy of system is closed relate to the flux of the cell cycle. This indicates the origin and preservation of the flux comes from the input of energy.

\begin{figure}[!ht]
\centering
\includegraphics[width=1.0\textwidth]{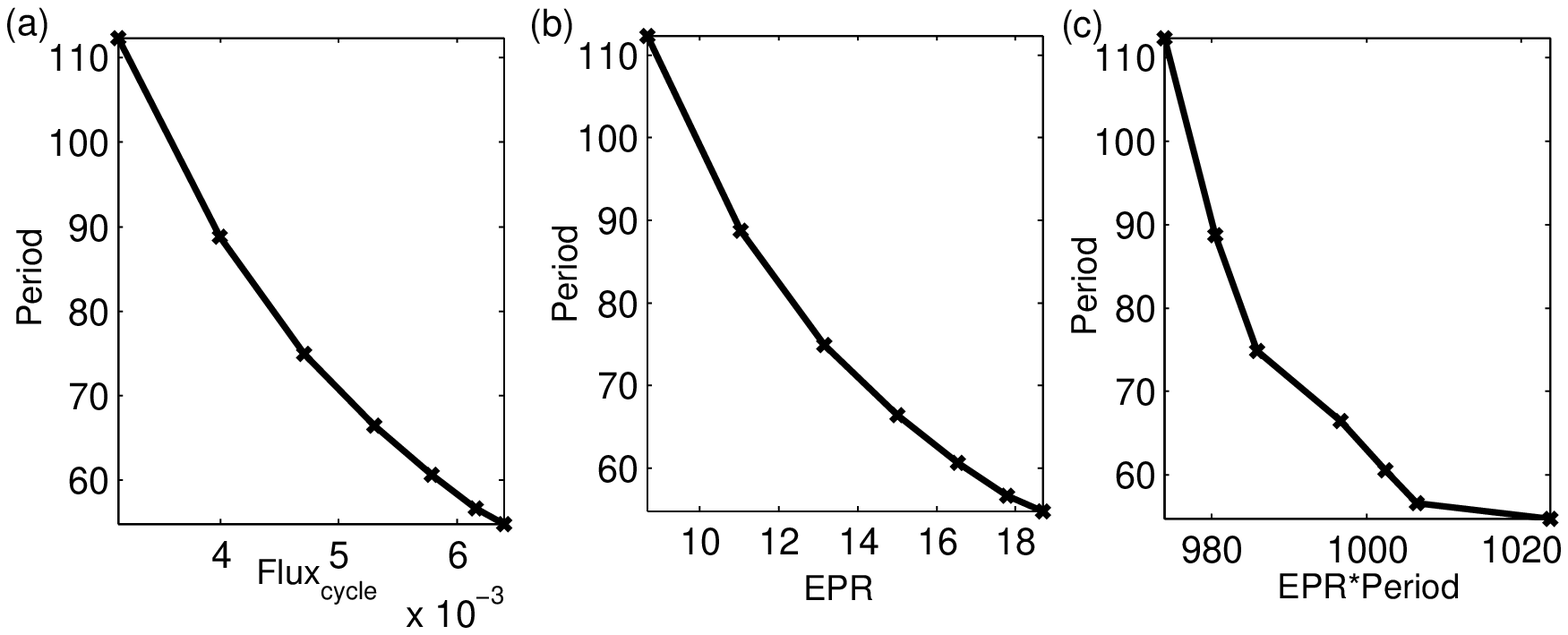}
\caption{(a)Entropy production rate with different period. (b)Entropy production per cycle with different period. (c)The integral of flux along the limit cycle with different period.}\label{fig12}
\end{figure}

Figure \ref{fig12}(a) shows that the period of cell cycle decreases with the increases of the flux. We found that the larger curl flux leads to the faster cell cycle oscillation. In Figure \ref{fig12}(b) and (c), we show the cell cycle period versus EPR and the energy cost per cycle. We see that the cell cycle period monotonically decreases with the increase of both the EPR and the energy cost per cell cycle. The faster(smaller period) the cycle oscillation of the cell growth and division is, the more the energy per cycle consumes. This indicates that the sufficient supply of the nutrition or energy pump is necessary to drive and accelerate the replication such as cell cycle.

\subsection*{Global Sensitivity Analysis of Key Genes and Regulations for Cell Cycle}
The process of the cell cycle is controlled by the interactions among many genes and gene regulations. To find the key genes and regulatory wirings in the cell cycle network, we perform a global sensitivity analysis of the cell cycle period, the flux integrated along the cell cycle path, EPR and the landscape barrier upon the moderate changes of the genes and wirings. By such global sensitivity analysis, we can identify the key structure elements (genes or gene regulations) or hot spots for the cell cycle network.

\begin{figure}[!ht]
\centering
\includegraphics[width=1.0\textwidth]{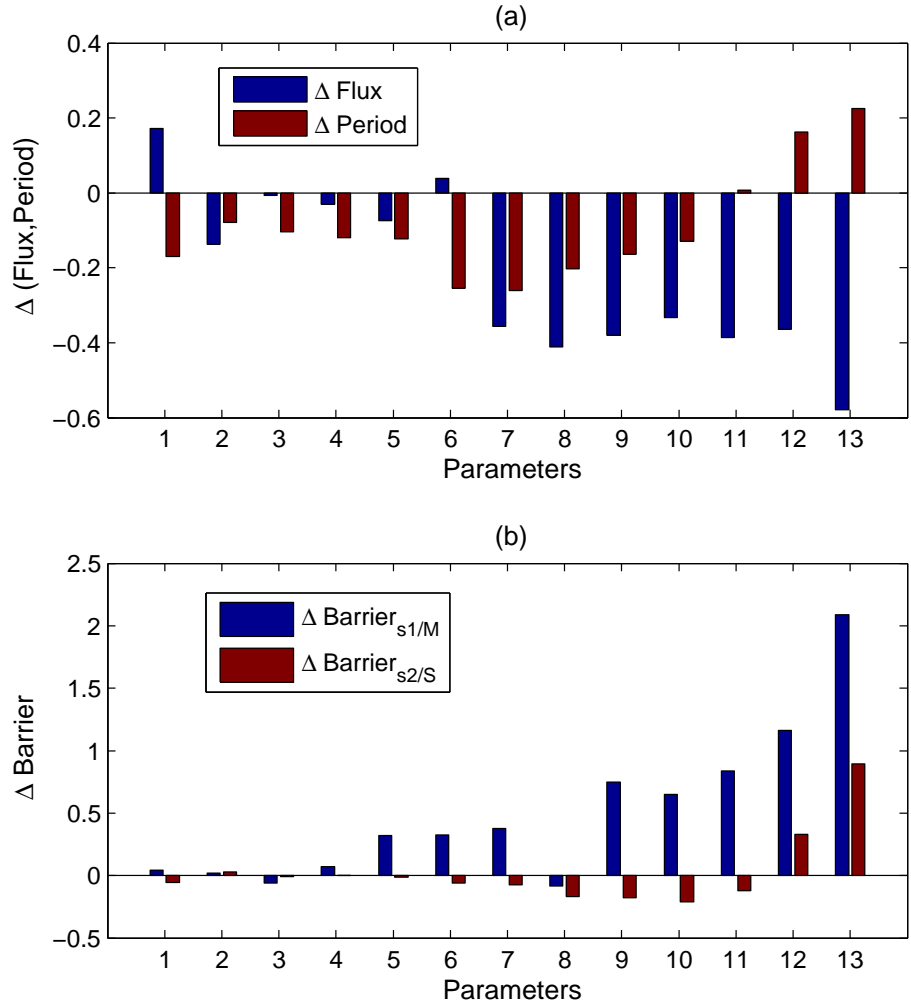}
\caption{(a)Global sensitivity analysis in term of the flux and period when parameters are changed. (b)Global sensitivity analysis in term of the barrier (Barrier between s1 and M, Barrier between s2 and S) when parameters are changed.The x coordinate(1 to 13) is corresponding to the 13 parameters. 1: $k_s$, 2: $a_{deg}$, 3: $b_{deg}$, 4: $EC50_{deg}$, 5: $n_{deg}$, 6: $a_{cdc}$, 7: $b_{cdc}$, 8: $EC50_{cdc}$, 9: $n_{cdc}$, 10: $a_{weel}$, 11: $b_{weel}$, 12: $EC50_{weel}$, 13: $n_{weel}$.}\label{fig13}
\end{figure}

Exploring important structural elements of the network (synthesis rate or regulation strengths), we show the results of the global sensitivity analysis in Figure 11. We analyzed the global sensitivity of the flux, EPR, the period and barrier upon changes of the 13 parameters in the the Xenopus laevis embryonic cell cycle network. We found certain key parameters in the cell cycle network from figure \ref{fig13}. They are $k_s$, $b_{cdc}$, $EC50_{cdc}$, $EC50_{weel}$ and $n_{weel}$. $k_s$ represents the synthesis rate of the cyclin. $b_{cdc}$ denotes the maximum rate of phosphorylation of Cdc25C by Cdk1. $EC50_{cdc}$ is the half-maximum effective concentration value of Cdc25C and denotes the concentration demanded of the substrate Cdk1 when Cdc25C achieves half of the activity. $EC50_{weel}$ is the half-maximum effective concentration value of Wee1A. $n_{weel}$ is the Hill coefficient. It represents the cooperative effect of the Wee1A regulated by Cdk1 and leads to the rapid decrease of the production rate with the increase of the substrate Cdk1.

Figure \ref{fig13}(a) shows the global sensitivity of the flux along the cell cycle path and the period upon parameter changes. We can see that certain key parameters have more significant effects on the flux or oscillation period. Therefore, the network can have larger fluxes and accelerate more of the cell division by adjusting these key parameters. For example, the B-type cyclines is the most important protein in mitosis, its synthesis rate is directly related to the progression of the cell cycle. Therefore, the synthesis rate parameter $k_s$ is one of the key parameters. From the figure \ref{fig13}(a), we can also see the increase of the rate can significantly decrease the period and increase the flux. Cdc25C is one of the protein kinases to be phosphorylated by Cdk1. It can re-activate the cyclin-Cdk1 complexes and then accelerate the cell cycle. We also found that the increase of the related maximum synthesis rate $b_{cdc}$ can significantly decrease the period. Wee1A is an early substrate of Cdk1 and inactivates the cyclin-Cdk1 complexes. Therefore, it can repress the cell cycle. Form the global sensitivity analysis, we can also see the increase of the related parameters 12 and 13 of Wee1A can increase the period and decrease the flux.

Figure \ref{fig13}(b) shows the change of the landscape barrier upon parameter change. The barrier$_{s1/M}$ and barrier$_{s2/S}$ respectively characterize the M check point and G1 check point. Therefore, the network can increase(or decrease) the check point barrier by the adjusting certain key parameters. For example, Cdc25C proteins are known to control the cell progression from G1 to S phase and G2 to M phase from biologic studies. Our analysis shows that the increase of the related maximum synthesis rate $b_{cdc}$ and the half-maximum effective concentration value $EC50_{cdc}$ of Cdc25C can change the activity of Cdc25C and decrease the barrier between s2 and $S/G2$. This implies that the increase of these regulation parameters can help the cell to go over the G2 checkpoint and accelerate the cell cycle. Wee1A is a key regulator of cell cycle progression and a component of a cell size checkpoint. It can inhibit the entry into mitosis. The related parameter $EC50_{weel}$ and $n_{weel}$ can significantly influence the activity of Wee1A. From the figure, we see that it increases the barrier between $s2$ and $S/G2$, and increases the barrier between $s1$ and $M$. Therefore, this can slow down the cell cycle. The above analysis on the key genes and regulations is consistent with the findings of Xenopus laevis embryonic cell cycle studies. Furthermore, the current approach can provide predictions for further experimental test.

\section*{Materials and Methods}
The dynamics of the gene regulatory network can often be described by a set of ordinary different equations. The cell cycle control dynamics realized by the underlying gene regulatory network can be described by ODEs. However, the deterministic description is not complete for the fluctuating environments of the gene regulatory networks. The intrinsic statistical fluctuations from the finite number of molecules inside of the cell and external fluctuations from cellular environments have significant impacts on the network dynamics. Therefore it is necessary for the dynamics of the regulatory network to be formulated as the stochastic different equations with the noise $\frac{\ud \bf x}{\ud t}=\bf F (\bf x) + \eta$, where $\bf x$ is the concentration or expression levels of the substance, $\bf F(\bf x)$ is the driving force of the system. $\eta$ is Gaussian white noise term with zero mean and its autocorrelation function is given as $<\eta(t)\eta(0)> = 2D \delta(t)$. $D$ is the diffusion coefficient. This characterizes the intensity of the intrinsic and cellular environmental fluctuations. the process is similar to Brownian dynamics and diffusion coefficient can be dependent on the concentration.

The time evolution of the expression or concentration dynamics is not deterministic because of the stochastic nature. More appropriate quantitative description can be obtained by the probability distribution. One can do statistical analysis and calculate out the probability distribution at stead state from the simulated trajectories of the underlying stochastic dynamics. On the other hand, the probability evolution follows the diffusion equations for continuous case\cite{Van1992-Stochastic}. The equation is also called Fokker-Plank equation which can be written in the form of probability conservation:$\partial P/\partial t + \nabla \cdot J=0$ where J is defined as the probability flux $J=FP -D \nabla P$. The equation states that the increase or decrease of the local probability is equal to the net input flux. When the divergence of the probability flux $\mathbf{J}_{ss}$ is zero($\nabla \cdot J=0$), the non-equilibrium system attains the steady state. We can solve out the steady state probability distribution $P_{ss}$ from the Fokker-Plank equation. We define the potential $U=-lnP_{ss}$ which resembles the Boltzmann law under equilibrium condition. If the local flux is equal to zero, then the detailed balance condition is satisfied and the system is in equilibrium state. When the local flux is not equal to zero, the detailed balance is broken and the system is in non-equilibrium steady state, we see that $\mathbf{F} = - D \cdot \nabla U + \mathbf{J}_{ss}/P_{ss}$. Thus, we have decomposed the force driving the dynamics of the system into two terms, The first is related to the gradient of the potential $U$ , the second term is the steady state probability flux $\mathbf{J}_{ss}$ (velocity current) divided by the steady-state probability $P_{ss}$ (density). The steady state flux is divergent free at steady state and therefore rotational termed as curl flux. The nonequilibrium dynamics is analogous to a moving electron in the electric and magnetic field.

The non-equilibrium system is an open system with exchanges in energy, materials and information to the environments. The system will generate energy consumption and dissipation. The dissipation as a global physical characteristic can be used to measure the degree of the non-equilibriumness away from the equilibrium.  The energy dissipation is associated with the entropy production rate in the steady state of non-equilibrium system \cite{Feng2011-234511,Qian2001-016102}. The system entropy can be written as $S=-\int P(\bf x, t) ln P(\bf x,t)d \bf x$. By differentiating the above expression, the change rate of the system entropy can be formulated as follow,
$\dot S =\int( \bf J \cdot \bf D^{-1} \cdot \bf J)/P d \bf x -
\int (\bf J \cdot \bf D^{-1} \cdot (\bf F -\nabla \cdot \bf D)) d
\bf x$,
where $\int( \bf J \cdot \bf D^{-1} \cdot \bf J)/P d \bf x= e_p =\dot S_{tot}$ is the entropy production rate(EPR). It represents the total entropy change rate (including both system and environment). $\int (\bf J \cdot \bf D^{-1} \cdot (\bf F -\nabla \cdot \bf D)) d \bf x = h_d=\dot S_{env}$ is the rate of the heat dissipation or the entropy change rate from the environment. When the non-equilibrium system is in a steady state, the change rate of the system entropy $\dot S$ is equal zero. Therefore, The entropy production rate is equal to the heat dissipation from the environment in steady state. According to the equation, the energy dissipation quantified by entropy production $e_p$ and $h_d$ is associated directly with the curl flux $\bf J$. The equation can also be written as $\dot S_{tot}=\dot S + \dot S_{env}$. This gives the first law of non-equilibrium thermodynamics. The entropy production is always larger or equal to zero. This gives the second law of non-equilibrium thermodynamics \cite{Feng2011-234511}.

\section*{Conclusions}
In this study, we explored the underlying mechanisms of the Xenopus laevis embryonic cell cycle through uncovering the underlying landscape and flux. Firstly, We quantified the underlying the landscape of the cell cycle controlled by the gene regulatory network. The potential landscape has an irregular Mexican hat shape. Secondly, we uncovered the relationship between the different phases of Xenopus laevis embryonic cell cycle and the landscape basins on the cycle. We identified the locations and quantified the potential barriers along the oscillation ring as the checkpoints of the Xenopus laevis embryonic cell cycle. This provides a physical quantification of the checkpoint mechanism of Xenopus laevis embryonic cell cycle. Thirdly, we uncovered the driving forces for the dynamics of Xenopus laevis embryonic cell cycle, the underlying landscape and the curl flux which measures the degree of detailed balance breaking. While landscape leads to the stability of the states on the Xenopus laevis embryonic cell cycle, the curl flux drives the persistent oscillation of the Xenopus laevis embryonic cell cycle. The potential barriers separate the oscillation into different phases and impede the progression of cell cycle. Finally, we want to emphasize that the replication is fundamental for biology of living. The cell replication requires the cost the certain energies to initiate and sustain. The curl flux originated from the nutrition supply and the corresponding energy consumption drive and complete the cell cycle process. Through landscape and flux analysis, we can identify several key elements for controlling the cell cycle speed. This can help to design effective strategy for drug discovery against cancer.
.
\section*{Acknowledgement}

We thanks the support of National Nature Science Foundation of China Grants No.91430217 and Most, China, Grant No.2016YFA0203200. JW thanks supports in part by NSF-PHY-76066.

\newpage



\newpage

\end{document}